# *Probing formation and epitaxy of ultrathin Titanium Silicide using low and medium energy ion scattering*

Philipp M. Wolf[1], Eduardo Pitthan[1], Zhen Zhang[2], Tuan T. Tran[1], Radek Holeňák[1], and Daniel Primetzhofer[1,3]

[1]*Department of Physics and Astronomy, Uppsala University, Uppsala, Sweden*

[2]*Division of Solid-State Electronics, Department of Electrical Engineering, Uppsala University, Uppsala, Sweden*

[3]*Tandem Laboratory, Uppsala University, Uppsala, Sweden*

*Corresponding Author: philipp.wolf@angstrom.uu.se*

**Abstract**

Titanium silicide is a key contact material in advanced three-dimensional semiconductor device architectures. Here, we examine the formation of ultrathin Ti-silicide on Si(100) using a combination of non-destructive *in-situ* and *ex-situ* ion scattering techniques capable of resolving composition and structure at the nanoscale. *In-situ* Time-of-Flight Low-Energy Ion Scattering (ToF-LEIS) indicates intermixing after annealing at 350 °C, with further compositional changes after annealing at 500 °C, including the emergence of a Si terminating layer at the surface. Consecutive *ex-situ* Time-of-Flight Medium-Energy Ion Scattering (ToF-MEIS) reveals a Ti-rich polycrystalline surface layer and a Si-rich interface layer exhibiting strong ordering along the Si [100] axis. High-Resolution Transmission Electron Microscopy (HR-TEM) images confirm these findings, revealing a ≈1.5 nm thick epitaxial silicide layer at the interface. The presence of an epitaxial interface is particularly promising for minimizing contact resistivity in ultrathin contact layers, where interfacial order can dominate electronic performance. In addition, both ToF-MEIS and HR-TEM unveil significant variations in the thickness of the silicide layer, with a substantial interface roughness but no translation of this roughness to the surface.

# 1. Introduction

Transition metal silicides experience widespread use as contact materials in integrated circuits due to their low electrical resistance and high metallurgical stability [1]. Over the last decades, multiple transition metal silicides have been used as contact materials, among them in order of reducing technology node size $C54-TiSi_2$, $CoSi_2$, and NiSi/NiPtSi [1]. With the advent of 3d device structures with extremely scaled dimensions, like FinFETs, Ti-based silicides experienced a renaissance. The reasons for this shift back to Ti silicides are an increase in the importance of the specific contact resistivity of the silicide/Si contacts compared to the sheet resistance of the silicide, originating from the design architecture [2], as well as the problem of Ni encroachment in 3d devices. While a shift back to Ti silicide does not lead to a large increase in specific contact resistivity, it overcomes the challenge of encroachment due to the higher melting point of $TiSi_x$ compared to NiSi, which implies a lower atomic mobility of Ti than Ni [2–5]. The high relevance of transition metal silicides in the semiconductor industry is attested by the detailed knowledge on their properties, like growth, phase transitions, electrical performance, et cetera collected in the last decades.

In the case of Ti silicides, a meta-stable $C49-TiSi_2$ with a $ZrSi_2$ structure and high resistivity is formed at 600 °C before the stable orthorhombic $C54-TiSi_2$, desirable due to its low resistivity, is formed above 800 °C [6,7]. At temperatures below 600 °C, multiple phases have been observed simultaneously, including alloyed Ti, amorphous silicide, $Ti_5Si_3$, and $C49-TiSi_2$ [8].

For the stable orthorhombic $C54-TiSi_2$ phase, multiple axiotaxial and epitaxial alignments have been observed on both Si (100) and Si (111) substrates [9–13]. While the C49 phase is reported to be predominantly oriented along the (100) planes of the substrate, the C54 phase is reported to be aligned with the (111) planes of the Si substrate [12,14].

While the above-described transition pathway of Ti silicide is well studied, most of the described effects have been observed on comparatively thick films with initial transition metal film thicknesses in the tens of nm. The continued miniaturization of integrated circuits leads to a decrease of silicide thickness into a realm in which interface effects towards the Si substrate can lead to significant changes in phase transition sequences of transition metal silicide [15–17]. At these reduced thicknesses, conventional reciprocal space techniques face challenges in obtaining reliable structural information. Due to these possible differences in ultrathin silicide films and the limitations of reciprocal space techniques, we aim to investigate the silicide formation for ultrathin Ti silicide using depth-resolved and non-destructive

medium- and low-energy ion scattering techniques supported by Auger electron spectroscopy, Low-energy electron diffraction and High-Resolution Transmission Electron Microscopy.

Medium- and Low-energy ion scattering (MEIS & LEIS) techniques provide an improved depth resolution at the cost of a reduction in information depth compared to more common ion scattering methods like Rutherford Backscattering Spectrometry (RBS) while maintaining their non-destructive nature, making MEIS & LEIS especially advantageous for analyzing ultrathin films, such as those investigated in this study. Beyond compositional analysis, ion scattering methods also can provide valuable real-space information on the morphology of thin films. By employing position-sensitive detectors, as done in the ToF-MEIS experiment presented in this work, it is possible to enhance morphological insights even further, capturing two-dimensional blocking patterns without the need for extensive angular scanning. The full power of the low-energy ion scattering experiments is harnessed by performing experiments with even higher surface sensitivity *in-situ*, thus avoiding the effects of reactions between atmospheric gasses and the sample surface.

We utilize two ion scattering approaches to study an ultrathin Ti silicide layer. First, we perform *in-situ* ToF-LEIS measurements to study the Ti silicide formation after annealing at 350 °C and 500 °C. Second, we perform *ex-situ* ToF-MEIS measurements to further investigate the structure and composition of the silicide layer after annealing at 500 °C. Lastly, we augment the ion scattering experiments with HR-TEM images. We chose the moderate annealing temperature of 500 °C, since studies have shown that the Ti silicidation can be performed at temperatures below 600 °C [3,18,19].

Our ion scattering experiments, presented in the following, reveal the presence of a highly ordered interface silicide layer as well as a high variation in silicide thickness after annealing at 500 °C. Both these findings are supported by the HR-TEM images.

## 2. Experimental Methods

A clean Si surface was prepared by multiple cycles of ion sputtering, using 3 keV $Ar^+$ at an incidence angle of 60°, and subsequent annealing at approximately 800 °C of a p-type Si (100) substrate (Silicon Materials, 10-20 Ω cm). Subsequent Auger Electron Spectroscopy (AES), Low-Energy Electron Diffraction (LEED), and ToF-LEIS measurements confirmed both the absence of surface contaminations and the expected Si(100) surface with a 2×1 surface reconstruction.

Following the substrate preparation, Ti was deposited on the Si (100) surface from a Ti-rod (MaTeck, purity 99.995%) using an $e^-$-beam evaporator (UHV Evaporator 3T by Omicron) at a voltage of 1 kV and a filament current of 2.25 A. During the Ti deposition, the pressure was continuously lower than $8.4 \times 10^{-7}$ Pa. The deposition time was set to 25 min, resulting in an expected Ti layer thickness of 3 nm. *Ex-situ* RBS measurements using a 2 MeV $He^+$ beam at the 5 MV pelletron accelerator at the Tandem Laboratory, Uppsala [20], performed after the completion of all *in-situ* measurements, revealed the actual areal density of the Ti film to be $21.3 \times 10^{15}$ atoms $cm^{-2}$ which corresponds to a thickness of 3.7 nm, assuming bulk density (4.52 g $cm^{-3}$).

Following the deposition and the subsequent AES and ToF-LEIS measurements, the sample was annealed at 350 °C and 500 °C for 10 min, respectively. After both annealing steps, during which the pressure did not rise above $1.5 \times 10^{-6}$ Pa, the ToF-LEIS and the AES measurements were repeated. All AES measurements were performed with a PHI 10-155 AES system utilizing a 3 keV beam energy. Relative concentrations of different elements were calculated based on the AES spectra according to Ref. [21]. The ToF-LEIS measurements were executed using a 9 keV $He^+$ beam at a scattering angle of 129° and a detector acceptance angle of 0.92° in the ACOLISSA setup at Uppsala University [20,22]. For better comparability of the ToF-LEIS spectra recorded before and after Ti deposition, as well as after the annealing steps, all ToF-LEIS spectra have been normalized to the yield below 1899 eV in the spectrum recorded after Ti deposition. In addition to spectra recorded at an incidence angle of 0° and an exit angle of 51°, angular scans, varying the incidence angle between -15° and 66° in steps of 3°, were performed, to gather information on the crystallographic structure of the silicide film [23].

ToF-LEIS measurements were further evaluated by comparing them to simulations performed with the TRBS code, a Monte-Carlo simulation code based on the TRIM code [24,25]. All TRBS simulations performed within this study treated the screening and scattering potentials according to the Ziegler-Biersack-Littmark (ZBL) potential [26]. Electronic stopping cross-sections within the simulations were based on Ref. [27] for Ti and on Ref. [28] for Si. In the case of titanium silicides, the electronic stopping cross section was approximated using Bragg's rule. To account for the finite and energy-dependent detector resolution of the actual experiments, the results of the TRBS simulations were convoluted with a Gaussian distribution with a full width at half maximum proportional to $E^{3/2}$. All TRBS simulations were performed with $3\times10^7$ projectiles and a cut-off angle of 2.3°, small enough to ensure a realistic effect of multiple scattering.

To complement the *in-situ* AES and ToF-LEIS measurements, *ex-situ* ToF-MEIS, and HR-TEM measurements were employed. ToF-MEIS measurements were conducted at the 350 kV ion implanter at the Tandem Laboratory, Uppsala University [20,29,30]. An electrostatically pulsed [31] primary beam of 50 keV $He^+$ ions was used for two distinct types of ToF-MEIS measurements: the first, in a so-called random orientation (scattering angle: 140°, incidence angle: 15°, exit angle: 25°), was aimed at acquiring a backscattering spectrum unaffected by crystallographic effects. The second measurement, at the same scattering angle but with an incidence angle of 40° and exit angle of 0°, was designed to record a real-space blocking pattern and thus probe the crystallographic structure of the sample. The resulting ToF-MEIS data were analyzed using CoboldPC (Computer-Based Online Offline List Mode Data Analyzer).

The ToF-MEIS backscattering spectrum was further evaluated using SIMNRA simulations [32], assuming the ZBL screened potential, multiple scattering according to the Szilagyi model, and dual scattering. The electronic stopping power was based on SRIM [33] with a correction factor of 1.19 for Si, based on a comparison between SRIM and experimental data at 50 keV [28], and a correction factor of 1.20 for Ti, based on a comparison between SRIM and experimental data at lower energies [27]. Interface roughness was simulated using a gamma distribution, and the simulation was normalized to the experimental counts in the Ti-peak region since the amount of Ti is known from the RBS measurement.

Lamellae for the HR-TEM measurements were extracted using a focused ion beam (FIM/SEM235), and *in-situ* lift out, with the silicide layer being protected by a Pt layer deposited prior to the focused ion beam usage. HR-TEM images were recorded using a FEI Titan Themis 200 system at an acceleration voltage of 200 kV.

## 3. Results

### 3.1 *In-situ* measurements

Figure 1a, shows an energy-converted spectrum recorded using ToF-LEIS before the deposition of Ti, indicating a clean Si surface with a distinct surface peak. As shown in the same figure, after the deposition of 21.3×10$^{15}$ atoms cm$^{-2}$ of Ti, we observe two apparent changes in the ToF-LEIS spectrum, the first being the appearance of a Ti peak in an energy range of about 6000 eV to 7000 eV, the second being a shift of the Si signal edge from its original position towards lower energies. This shift of the Si signal edge to lower energies indicates a complete coverage of the Si substrate by Ti and speaks against large-scale intermixing during the Ti deposition. The finding that Ti indeed completely covers the whole Si surface is supported by our Auger Electron Spectroscopy (AES) measurements, Figure 1b, in which we do not observe a Si signal after the completed deposition. The relative surface concentrations based on this AES measurement are 77% Ti, 14% C and 9% O.

In addition to the ToF-LEIS spectra, Figure 1a includes multiple TRBS simulations, among them a simulation of 21.3×10$^{15}$ atoms cm$^{-2}$ of Ti on Si, the Ti amount indicated by our RBS measurements. Here, we observe a good agreement in the shape of the Ti peak with the ToF-LEIS experiment, while we see a significant discrepancy between the TRBS simulation and the ToF-LEIS experiment below 5400 eV, where the yield in the simulation is higher than in the experiment. The straightforward reason for this discrepancy between simulation and experiment is the single crystalline nature of the Si substrate, which, due to channeling effects, often leads to a reduction in the backscattering yield. Since the crystal structure cannot be included in a TRBS simulation and thus an amorphous Si is assumed, the backscattering yield found in the simulation is significantly higher than observed in the experiment. Nonetheless, the excellent agreement between simulation and experiment for the Ti-peak range indicates the presence of a poly-crystalline or amorphous Ti layer with a thickness of 21.3×10$^{15}$ atoms cm$^{-2}$ (3.7 nm, assuming bulk density) covering the Si substrate.

After annealing the sample at a temperature of 350 °C, we observe a noticeable change in the ToF-LEIS spectra presented in Figure 1a, with a broadening of the Ti peak and a reduction in its height. These changes in the Ti peak range clearly indicate intermixing between the Ti and the Si. Even though the TRBS simulations can only be of limited use in analyzing these ToF-LEIS spectra due to their disregard of the crystal structure, we do observe a relatively good agreement between the ToF-LEIS spectrum recorded after annealing at 350 °C and the TRBS simulation assuming a homogenous Ti$_5$Si$_3$ surface layer.

While we observe clear signs of intermixing due to the annealing at 350 °C in the ToF-LEIS spectrum, we do not observe a re-appearance of the Si signal in the AES spectrum recorded after the annealing at 350 °C (Figure 1b). We detect a decrease in the Ti signal intensity to a relative concentration of 48% Ti, alongside an increase in the O and C signals to relative concentrations of 18 % O and 34% C. These AES results indicate that not the whole amount of metallic Ti has reacted with the Si substrate, and a thin layer of Ti, possibly Ti oxide, remains at the sample surface.

Angular scans in ToF-LEIS were performed to gain information on the morphology of the silicide layer. The results of a polar scan on the sample after annealing at 350 °C are presented in Figure 2. Here, we do not observe an angular dependence of the yield in the Ti-peak (6190 eV to 7103 eV) apart from the general increase of the yield with the polar angle. Hence, no preferred orientation is observed in the silicide layer.

Annealing the sample at 500 °C leads to a more significant change in the ToF-LEIS spectrum, presented in Figure 1a, with a reduction of the Ti signal height by a factor of roughly two, compared to the measurement performed after annealing the sample at 350 °C. Alongside the change in the Ti-peak area, we observe increased counts at energies of 5000 eV to 5600 eV. Both effects indicate further intermixing; The Ti-peak reduces in height since a significant amount of the Ti atoms is now positioned deeper in the sample, which also leads to an overlap of the Ti signal, now forming a plateau, extending to lower energies, i.e. lower depth, and the Si signal and thus to an increase in counts between 5000 eV and 5600 eV. The absence of a surface peak in the signal of projectiles backscattered from Ti alongside the rise in the signal of projectiles backscattered from Si possibly indicates a Si termination of the surface. A comparison to TRBS simulations of different silicide phases shows no excellent agreement between the simulated silicide layers and the experimental spectrum. The best agreement is reached for $TiSi_2$, which agrees well with the experimental spectrum at the high energy edge of the Ti signal but overestimates the yield compared to the experiment for lower energies. Here again, a reason for the discrepancy might be the structure of the silicide and the substrate and the resulting channeling effects.

While the Si signal re-appears in the AES spectrum after annealing at 500 °C (Figure 1b), it remains faint, resulting in a relative concentration of Si of 17%. Both the C and the O signal experienced a further increase due to the second annealing with resulting relative concentrations of 31% C and 25% O. The intensity of the Ti signal decreased significantly, and the resulting relative concentration of Ti is 27%.

In the polar scan, performed after annealing at 500 °C, presented in Figure 2, we do not observe a strong dependence of the yield on the polar angle. While we do not observe strong effects of a possible preferred orientation in the silicide layer, we do observe two small variations in the angular scan, one at a polar angle of 0° with a local yield maximum and one at a polar angle of 40° with a local yield minimum. While both variations in the yield are small, they can indicate a preferred orientation in parts of the silicide layer.

**3.2 *Ex-situ* measurements**

In order to gain additional insights into the silicide layer formed after annealing at 500 °C, in particular on the crystal order of the layer, we performed additional *ex-situ* measurements, starting with ToF-MEIS. While ToF-MEIS generally offers a somewhat lower depth resolution than ToF-LEIS due to the higher projectile energy, it enables a better kinematic separation of the Ti peak from the signal of the Si substrate. The backscattering spectrum resulting from the ToF-MEIS measurement is included in Figure 3 and shows two significant characteristics.

The first is a step in the Ti-peak at an energy of roughly 36 keV, which indicates the presence of two different silicide phases in the layer. A higher Ti signal for energies above 36 keV implies a Ti-rich surface phase, while the reduced Ti signal below 36 keV suggests a Si-rich interface phase.

The second significant observation in the ToF-MEIS spectrum is the low energy tail of the Ti-signal, which is less steep than expected for a homogenous Ti-silicide layer on top of Si. This flattened, low-energy edge of the Ti-peak strongly indicates a non-constant silicide layer thickness or, in other words, an interface roughness between the silicide and the Si.

These observations are supported by the SIMNRA simulations shown in Figure 3. The simulations, which account for a total Ti amount of $21.3 \times 10^{15}$ atoms cm$^{-2}$, model the layer as consisting of two distinct silicide phases: a surface layer containing $7.5 \times 10^{15}$ atoms cm$^{-2}$ of Ti$_2$Si and an interface layer with $48.9 \times 10^{15}$ atoms cm$^{-2}$ of TiSi$_2$. To match the experimental data, the model includes an interface roughness between the silicide and the Si substrate, with a full width at half maximum of $42.0 \times 10^{15}$ atoms cm$^{-2}$. The figure also shows a simulation without interface roughness, demonstrating that only by including this roughness can the flatter backside slope of the Ti peak be accurately explained. The missing intensity between 30 and 32.5 keV in either simulation can to largest extend be explained by the dual scattering model in SIMNRA, underestimating the true yield.

While the ToF-MEIS backscattering spectrum already provides us more insights into the silicide layer, we can gain additional information by making use of the position-sensitive detector in the ToF-MEIS setup. By mapping the backscattering yield over the position on the detector, we obtain an angular-resolved blocking pattern, which yields real-space information on the crystallographic structure of the sample, influencing backscattered ions on the way out of the sample which can induce blocking patterns. By filtering this map for different energies, we gain information on the crystallographic structure at different depths in the sample. Figure 4 includes three maps representing the backscattering yield from different depths in the sample.

For the surface region of the Ti silicide, the blocking pattern of which is presented in Figure 4a, we do not observe local yield minima, indicating the absence of long-range ordered crystal channels or planes. The absence of such effects of the crystal structure indicates a poly-crystalline or amorphous structure of the near-surface region of the silicide. The interface region shows a local yield minimum in the blocking pattern (Figure 4b) close to the center of the detector, indicating a preferred orientation of the silicide crystallites at the interface to the substrate around a common crystal axis, perpendicular to the sample surface. The position of the minimum observed for the silicide interface layer agrees with the minimum observed for the Si substrate, the blocking pattern of which is presented in Figure 4c. The agreement in position between the minima indicates that the silicide crystallites in the interface region are oriented along the Si [100] axis. In the blocking pattern obtained for energies corresponding to scattering from the Si substrate, we also observe a reduced yield along two crystal planes diagonally across the detector. The cut along the Y-axis through the blocking patterns, with −2 mm ≤ X ≤ 2 mm, presented in Figure 4d, supports our observation of the presence of a minimum at the same angular orientation for the Si substrate and the interface area of the silicide. The general increase of the backscattering yield from small to large X values is caused by the different scattering geometry, i.e., a decrease in scattering angle and, thus, increase in the scattering cross sections.

HR-TEM images, presented in Figure 5, agree with the ToF-MEIS findings. As indicated by the ToF-MEIS backscattering spectrum, we observe significant variations in the layer thickness in the HR-TEM image (Figure 5a), as well as the presence of two phases (Figure 5b & c). The HR-TEM images further show that the surface silicide layer is poly-crystalline, while the interface layer, visible in most of the silicide layer with a thickness of ≈1.5 nm, is epitaxial; findings which show an excellent agreement with the conclusions based on the ToF-MEIS blocking patterns. Additionally, we observed local extensions of the epitaxial interface layer to thicknesses of ≈5 nm in connection with phases showing axiotaxial orientation, examples of which are included in Figure 5c.

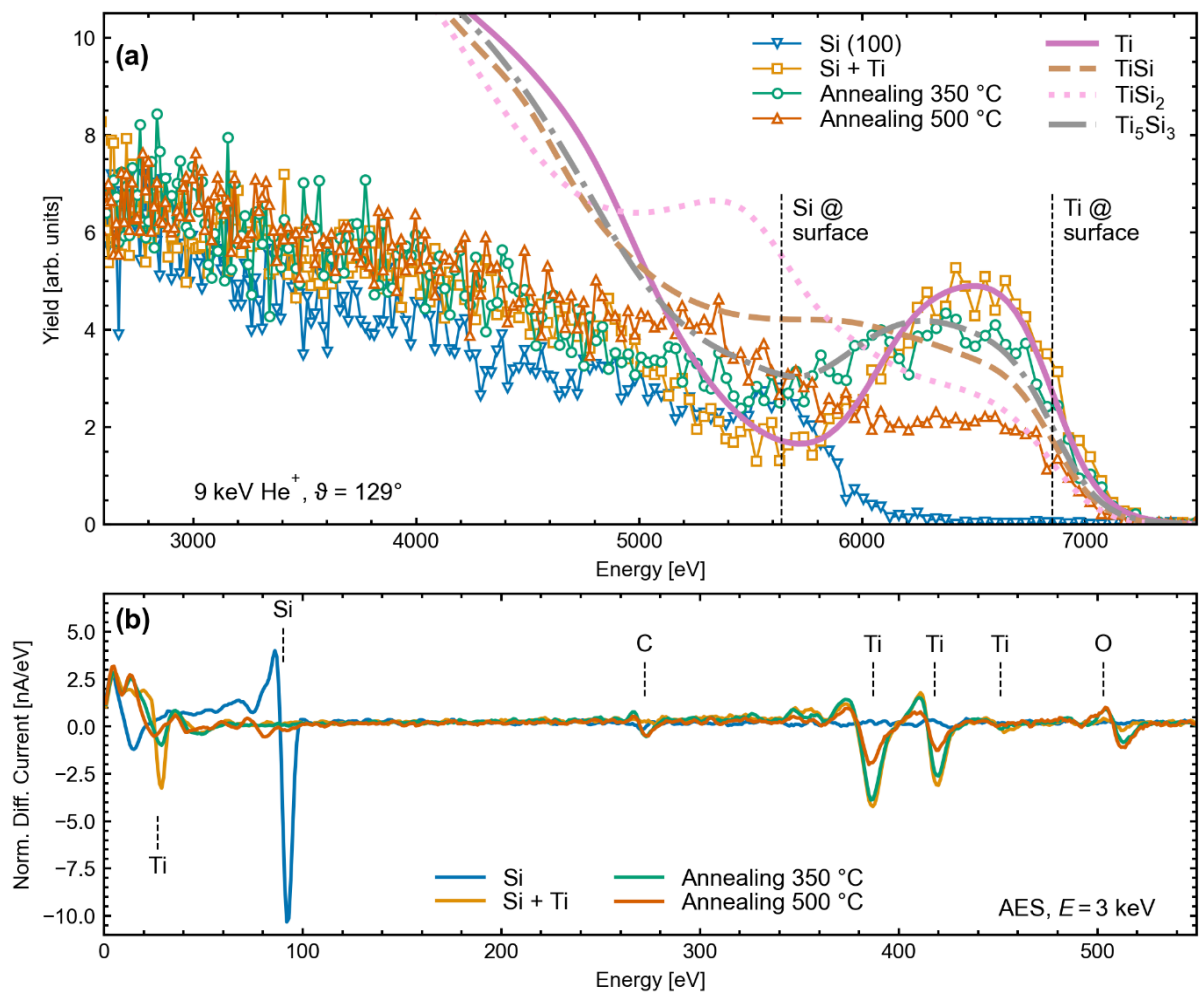

Figure 1: (a) Energy converted ToF-LEIS spectra recorded using 9 keV He$^+$ primary ions detected in a scattering angle of 129°. Data recorded for the clean Si (100) surface, after Ti deposition, after annealing at 350 °C, and after annealing at 500 °C are included. Multiple TRBS simulations, assuming different titanium silicide phases, are shown. (b) AES spectra recorded with a primary electron beam energy of 3 keV before the Ti deposition, after the Ti deposition, after annealing at 350 °C, and after annealing at 500 °C. Relevant elemental signals are marked.

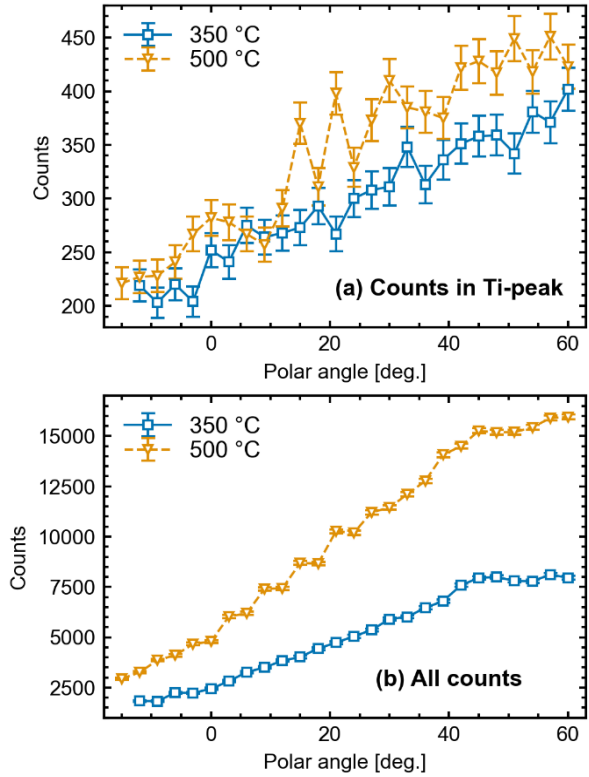

*Figure 2: (a) ToF-LEIS scattering yields in the energy range from 6190 eV to 7103 eV, corresponding to projectiles scattered from Ti as a function of the incidence polar angle, and (b) the total recorded scattering yield in each spectrum over the polar angle at which the spectrum was recorded. All spectra on which the angular scan is based were recorded using 9 keV He$^+$, a scattering angle of 129°, and an incidence angle of 0°.*

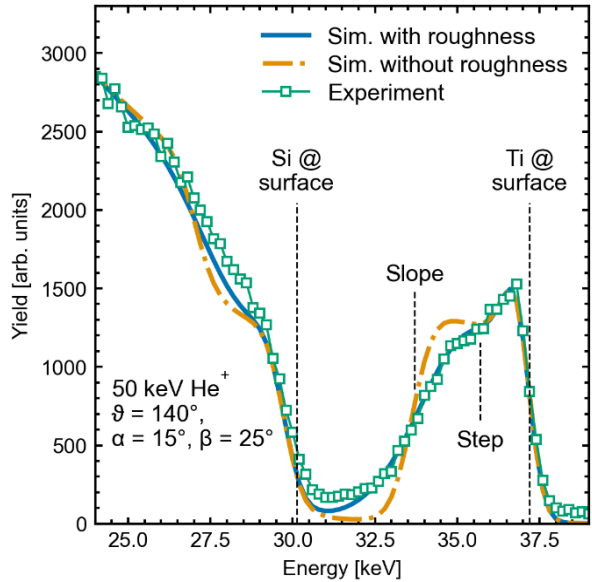

*Figure 3: Energy-converted ToF-MEIS spectrum of the sample after the final annealing step at 500 °C recorded with a beam of 50 keV He$^+$ primary ions detected in a scattering angle of 140°, for an incidence angle of 15°, and an exit angle of 25°. The figure marks two significant properties of the spectrum: a step in the Ti peak and the tilted low-energy edge of the Ti peak. Additionally, two*

SIMNRA simulations are included, one assuming an interface roughness between Si and silicide and one assuming no roughness.

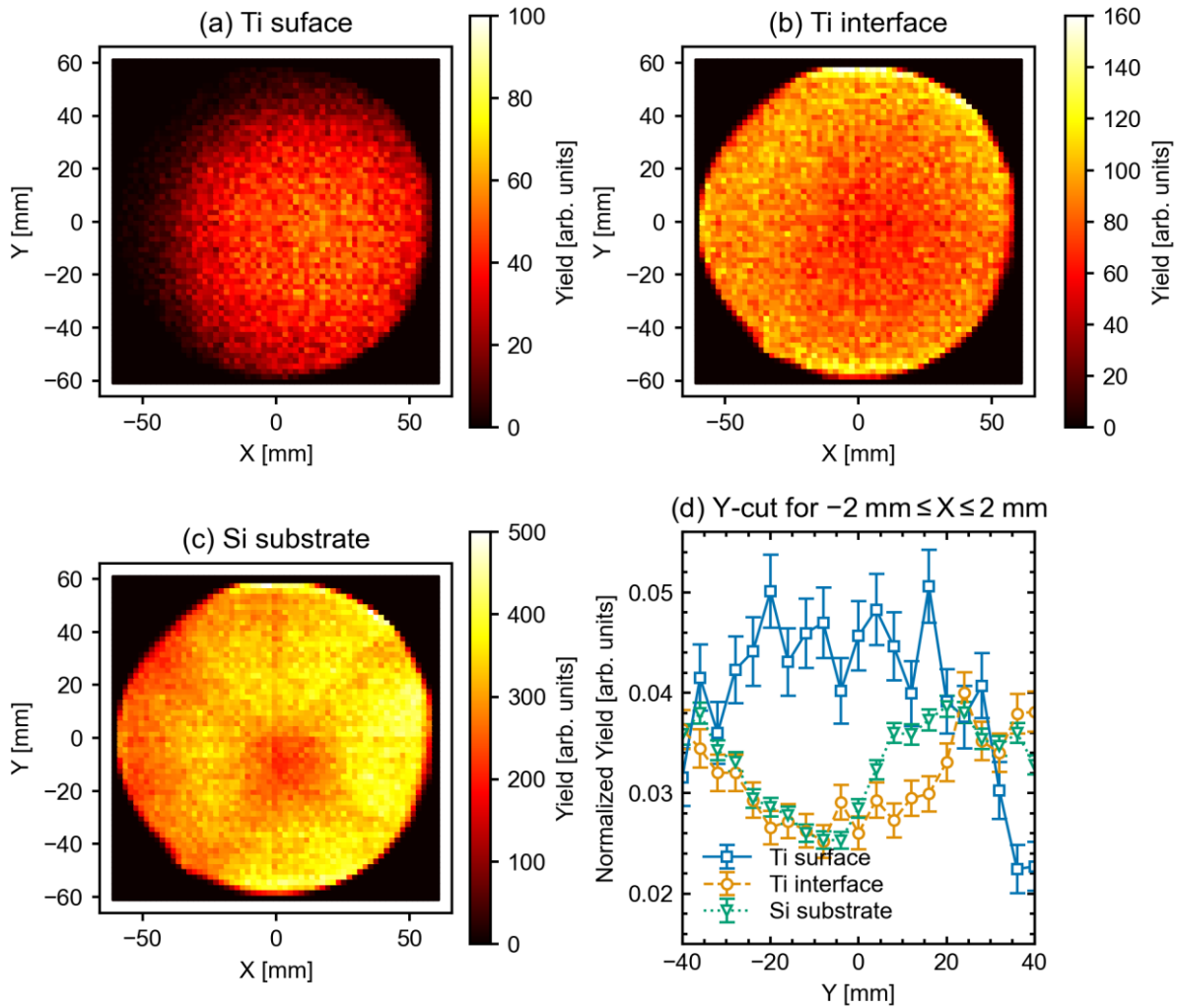

Figure 4: Blocking patterns recorded using ToF-MEIS with 50 keV He$^+$ primary ions, a scattering angle of 140°, an incidence angle of 40°, and an exit angle of 0°. (a) The position-dependent scattering yield of projectiles with a final energy between 35.94 keV and 38.37 keV representing projectiles scattered in the surface region of the Ti silicide, (b) for projectiles with a final energy between 31.44 keV and 35.94 keV, representing projectiles scattered in the interface silicide layer, (c) for projectiles with a final energy between 22.05 keV and 27.96 keV, representing projectiles scattered in the Si substrate, over the position on the detector. While no minimum is visible in the center of the detector in (a), such a minimum is visible in (b) and (c), indicating a preferred orientation of the interface layer along the Si [100] axis. The minima are also visible in a cut through the blocking patterns along the Y-axis, presented in (d).

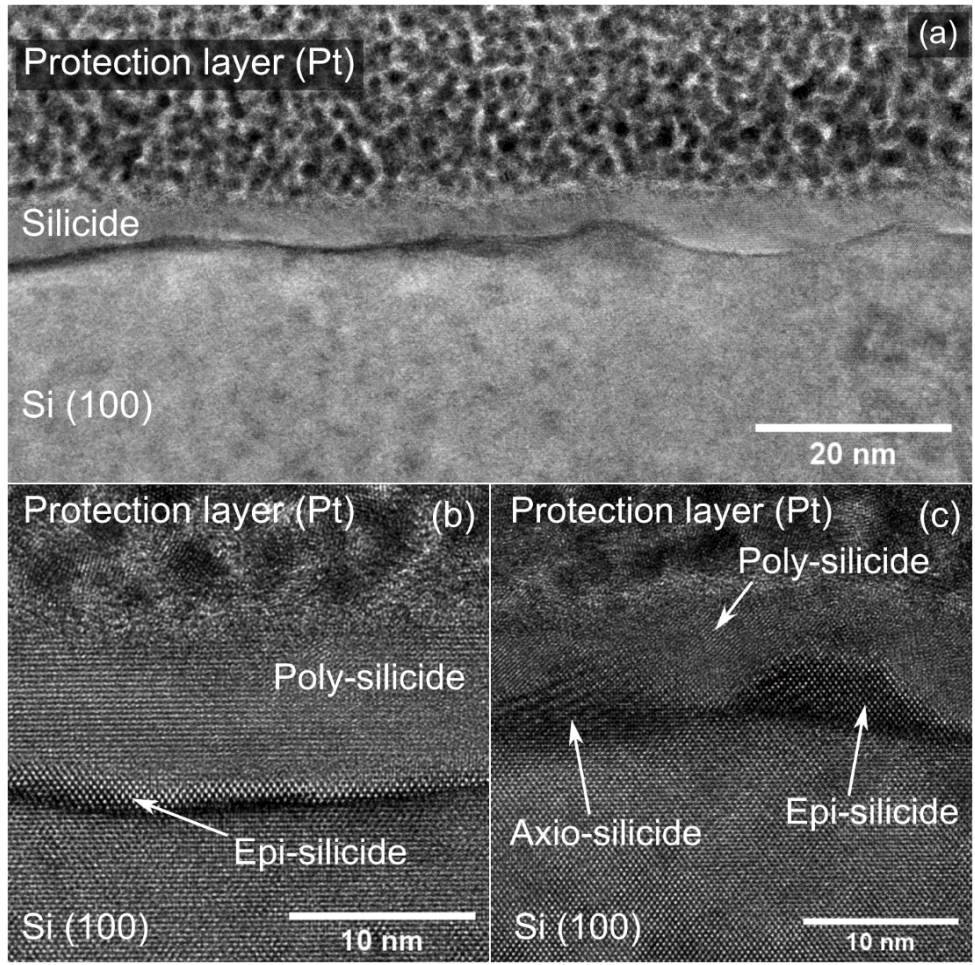

*Figure 5: HR-TEM images of the silicide layer after annealing at 500 °C. (a) A wide-field view of the silicide layer, showing significant variations in the thickness of the silicide layer. (b) A higher magnification image of the silicide showing the presence of a poly-crystalline surface layer and an epitaxial interface layer with a thickness of ≈ 1.5 nm. (c) Another high-resolution image of the silicide layer showing local extensions of the epitaxial interface layer to thicknesses of ≈ 5 nm and additional axiotaxial crystallites.*

## 4. Discussion

Previous studies reported an island growth of Ti and no significant intermixing for a clean Si (100) surface produced by sputter cleaning and subsequent annealing at 850 °C [34,35]. For a Si (100) surface prepared only by ion sputtering and thus presumably amorphized, a Stranski-Krastanov growth mechanism up to 4 ML of TiSi$_{0.78}$ and subsequent metallic Ti islands was reported [36]. These reports agree with our finding of a mostly metallic Ti layer on Si (100) after the deposition.

The observation of first signs of intermixing after annealing at 350 °C with ToF-LEIS agrees well with previous reports of the formation of an amorphous interface layer consisting of Ti and Si at annealing temperatures between 250 °C and 350 °C [37–40]. Different compositions have been reported for this amorphous layer, with compositions close to TiSi [37,41] and a composition gradient from close to TiSi$_2$ for the interface with the Si to a composition close to Ti$_2$Si at the interface to the metallic Ti [38]. According to Ogawa *et al.* and Cheng and Chan, the formation of the amorphous interlayer can be seen as a precursor to crystal formation and, subsequently, epitaxial silicide phases [38,42]. Our observation of an average composition close to Ti$_5$Si$_3$ would agree with previous studies that documented the presence of poly-crystalline Ti$_5$Si$_3$ on the interface of Ti on both Si (111) and Si (100) surfaces after annealing at 450 °C [40,43]. However, since the TRBS fit only gives an average composition over the whole layer, and we do not observe Si at the surface in AES, it is more likely that we have a gradient in the Si content from a higher concentration at the interface to a lower concentration closer to the surface.

Our observation of a highly ordered interface and a poly-crystalline surface after annealing the sample to 500 °C agrees with previous studies, in which the growth of C49-TiSi$_2$ crystallites has been observed to start between 450 °C and 500 °C [38,39,41]. Concerning the epitaxial interface layer observed in ToF-MEIS and HR-TEM, epitaxial C54-TiSi$_2$ layers on Si (100) have been reported in multiple studies although generally at temperatures higher than 500 °C [12–14]. Nevertheless, we do not expect the presence of C54-TiSi$_2$ here, on the one hand, due to the low annealing temperature of 500 °C, and on the other hand, due to the low film thickness of initially 3.7 nm Ti. For such thin layers, Jeon *et al.* [44] report that the C49 phase is no longer metastable but stable, and no transition to C54 occurs, even at high annealing temperatures. Thus, it is more likely that the epitaxial phase observed in our measurements after annealing at 500 °C is C49-TiSi$_2$, which has been reported to be oriented along the Si (001) surface planes [12].

Epitaxial interface layers of silicides on Si are known to potentially reduce the Schottky barrier height, thereby lowering the specific contact resistivity [45]. Consequently, the presence of such an epitaxial interface layer at the comparatively low temperature of 500 °C, as observed in this study, may significantly influence the performance of semiconductor devices. This is especially the case for 3d device architectures, in which the specific contact resistivity plays a larger role than the sheet resistivity [2].

Ti silicide is known to agglomerate in islands after annealing at high temperatures >800 °C [39]. Here, we observe roughening of the silicide layer at significantly lower temperatures.

## 5. Conclusions

In this study, the formation of ultrathin Ti-silicide layers on Si (100) was characterized through a combination of *in-situ* and *ex-situ* techniques. *In-situ* ToF-LEIS measurements provided direct evidence of intermixing after annealing at 350 °C. Upon further annealing at 500 °C, further compositional changes were observed, including the appearance of Si at the surface, a possible Si termination of the surface, and a likely evolution towards a $TiSi_2$ composition.

*Ex-situ* measurements using ToF-MEIS and HR-TEM revealed significant variation in the silicide layer's thickness and structure. ToF-MEIS spectra and blocking patterns indicate the presence of both a poly-crystalline, Ti-rich, surface silicide layer and a Si-rich interface layer highly oriented along the Si [100] axis. This observation was confirmed by HR-TEM images, which showed a ≈1.5 nm thick epitaxial silicide interface, contrasting with the poly-crystalline Ti-rich surface layer. The observed epitaxial silicide interface layer may play a critical role in reducing contact resistivity and enhancing semiconductor device performance.

These findings underscore the analytical power of non-destructive low-energy ion scattering techniques and *in-situ* measurements in resolving the complex morphology and compositional changes in ultrathin Ti-silicide films. The combination of high-depth resolution and real-space morphological information provided by ToF-LEIS, ToF-MEIS, and HR-TEM enabled a detailed resolution of both the epitaxial interface and the poly-crystalline surface layers, offering insights into the behavior of silicides at the nanoscale.

**Acknowledgments**

Financial support from the Swedish Research Council (VR) (contract # 2020-04754_3) and support of the acceleration operation at Uppsala University by VR-RFI (contract # 2019-00191) and the Swedish Foundation for Strategic Research (SSF) (contract # RIF14-0053) is gratefully acknowledged.